\documentclass[epjST]{svjour}
\usepackage{graphics}
\begin{document}
\title{The invisible Majorana bound state at the helical edge}
\author{Christoph Fleckenstein \and Felix Keidel \and Bj\"orn Trauzettel \and Niccol\'o Traverso Ziani }
\institute{Institute for Theoretical Physics and Astrophysics,
	University of W\"urzburg, Am Hubland, D-97074 W¨urzburg, Germany}
\abstract{The presence of a Majorana bound state in condensed matter systems is often associated to a zero bias peak in conductance measurements. Here, we analyze a system were this paradigm is violated. A Majorana bound state is always present at the interface between a quantum spin Hall system that is magnetically gapped and a quantum spin Hall system gapped by proximity induced s-wave superconductivity. However, the linear conductance could be either zero or non-zero and quantized depending on the energy and length scales of the barriers. The transition between the two values is reminiscent of the topological phase transition in proximitized spin-orbit coupled quantum wires in the presence of an applied magnetic field. We interpret the behavior of the conductance in terms of scattering states at both zero and non-zero energy.} %end of abstract
\maketitle
\section{Introduction}
\label{intro}
The metallic edges of two-dimensional topological insulators represent genuinely new gapless electronic systems\cite{ti1,ti2,ti3,ti4,ti5,ti6}. Their novelty lies in the fact that they can only exist as boundaries of higher dimensional structures\cite{ti6}. One of the most relevant characteristics making such states appealing from the point of view of applications is spin-momentum locking: electrons with opposite momentum have perpendicular spin projection\cite{ti6}. This behavior could have important implications in spintronics\cite{st1,st2,st3}, since it allows for purely electric manipulation of the spin degree of freedom. Moreover, when proximitized with s-wave superconductors and eventually ferromagnetic barriers, spin-momentum locked systems can develop topological\cite{sc1,sc2,sc3,sc4,sc41} and odd frequency\cite{sc5,sc6,sc7,sc8,sc9,sc10,sc11,sc12,sc13,sc14} superconductivity. This opens the way to potential applications in superconducting spintronics\cite{sst1,sst2}. Moreover, through the generation of Majorana zero energy bound states\cite{sc1,sc2} and, for strong electron-electron interactions\cite{ee1,ee2,ee3}, parafermions\cite{para3,para1,para2}, topological heterostructures can be useful in topological quantum computation\cite{tqc}. As far as the implementation of Majorana bound states is concerned, however, spin-orbit coupled quantum wires\cite{w0,w1,w2} in the presence of magnetic fields and induced superconductivity seem nowadays a promising host. They have attracted a huge amount of both experimental\cite{m1,m2,m3} and theoretical work\cite{m4,m5,m6}.\\
In this article, we show that the characteristics of Majorana fermions in spin-orbit coupled quantum wires and in two-dimensional topological insulators can drastically differ. While, in the first case, the scattering matrix formulation of the topological invariants\cite{smi1,smi2,smi3} shows that the local linear conductance is quantized to $2e^2/h$ whenever the Majorana bound state is present, in the case of heterostructures based on two-dimensional topological insulators such a correspondence is not valid. In particular, we consider the one dimensional metallic edge of a two dimensional topological insulator (a quantum spin-Hall system) partially proximitized by an s-wave superconductor and partially covered by a ferromagnetic barrier. We show that a Majorana bound state is always present at the interface between the two gapped regions, but the conductance can either be close to zero or to $2e^2/h$ depending on the parameters of the barriers. We interpret the results by analyzing the scattering states at both zero and finite energy.\\
The outline of the paper is the following one. In Sec.2 we introduce the model, in Sec.3 we demonstrate the presence of the Majorana bound state and we discuss the scattering states a zero energy. In Sec.4 we interpret the results on a more general basis and we consider finite bias. In Sec.5 we draw our conclusions.
\section{Model}
\label{sec:1}
We consider a quantum spin Hall system gapped by a superconducting barrier for $-L_s<x<0$ and by a ferromagnetic barrier for $0<x<L_D$. Explicitly, the Hamiltonian $H$ is given by
\begin{equation}
H=\frac{1}{2}\int_{-\infty}^\infty dx \Psi^\dag(x) \mathcal{H}(x) \Psi(x),
\end{equation} 
with the spinor given by $\Psi^\dag(x)=(\psi^\dag_+(x),\psi^\dag_-(x),\psi_-(x),-\psi_+(x))$, where $\psi_{\pm}(x)$ is the fermionic annihilation field operator for spin $+/-$ fermions and the Bogoliubov-de Gennes Hamiltonian reads
\begin{equation}
\mathcal{H}(x)=-i\partial_x\tau_z\sigma_z-\mu_0\tau_z\sigma_0+\Delta(x)\tau_x\sigma_0+M(x)\tau_0\sigma_x.
\end{equation} 
Here, $\mu_0$ is the chemical potential, $\Delta(x)=\Delta_0\theta(x+L_s)\theta(-x)$, $M(x)=M_0\theta(x)\theta(-x+L_D)$, with $M_0$ and $\Delta_0$ positive, and $\tau_i$ and $\sigma_i$ ($i=0,..,3$) are the Pauli matrices acting on particle-hole and spin sectors, respectively. Moreover, $\hbar=v_F=1$, with $v_F$ the Fermi velocity. For clarity, we will restore $\hbar$ and $v_F$ in the figures and in the main results. A schematic of the system is shown in Fig.1. It is worth to notice that a $\sigma_z$ component of the magnetization would not qualitatively affect the results.
\begin{figure}
	\resizebox{0.75\columnwidth}{!}{%
		\includegraphics{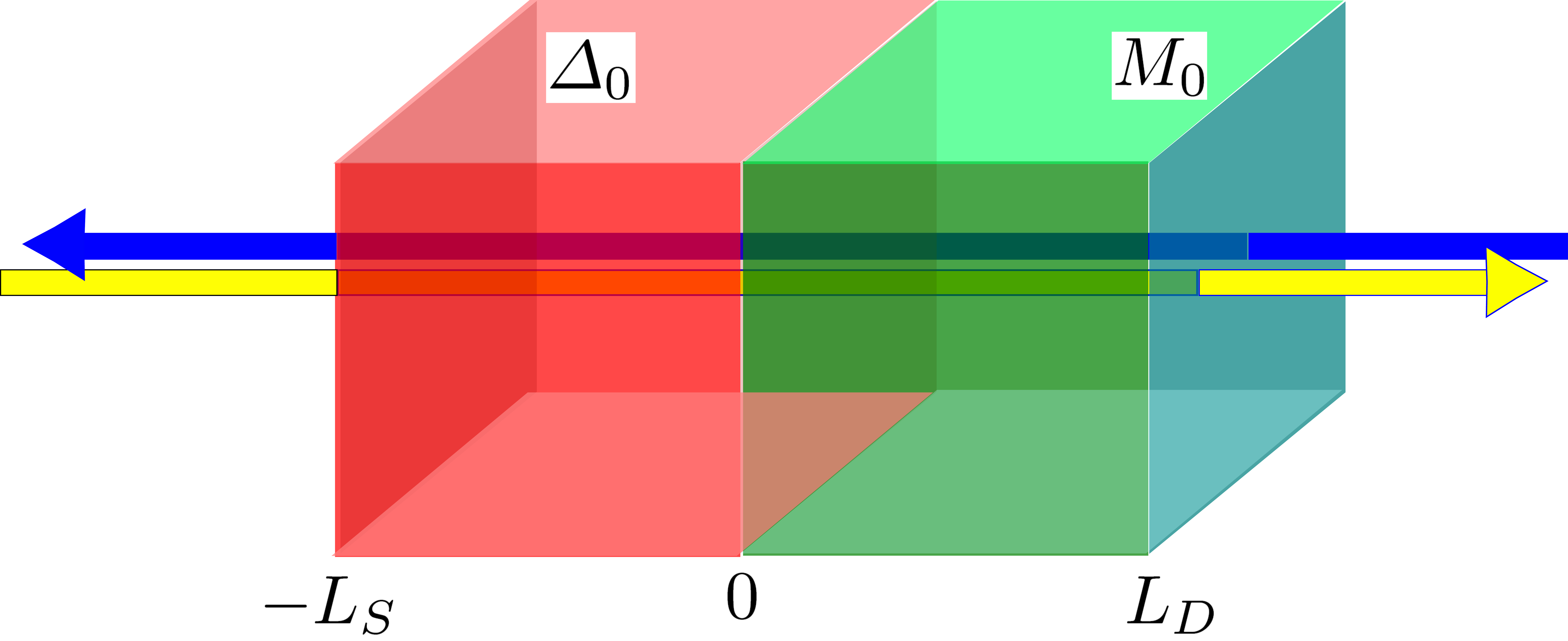} }
	\caption{Schematic of the system. The blue (upper) arrow indicates spin down particles and the yellow (lower) arrow indicates spin up particles. The red (left) cube indicates the superconductor while the green (right) cube indicates the magnetic barrier.}
	\label{fig:1}       % Give a unique label
\end{figure}
\section{Spectral and transport properties at zero energy}
\subsection{Zero energy decaying solutions}
We solve the eigenvalue problem $\mathcal{H}(x)r^T(x)=0$, with $r(x)=(\alpha(x),\beta(x),\gamma(x),\delta(x))$, that captures the zero energy states of the systems. Additionally, we require that the norm of the solution is maximal at $x=0$. For any set of non-zero parameters $\Delta_0$, $M_0$, $L_D$, $L_S$, and as long as $|\mu_0|<B$, a single solution can be found. Up to a real normalization constant, it reads, for $-L_S<x<0$,
\begin{equation}
r(x)=c_1\left(e^{(i\mu_0+\Delta_0)x},e^{i\varphi+(-i\mu_0+\Delta_0)x},i e^{(i\mu_0+\Delta_0)x},-i e^{i\varphi+(-i\mu_0+\Delta_0)x}\right)
\end{equation}
and, for $0<x<L_D$,
\begin{equation}
r(x)=c_1\left( 1,e^{i\varphi},-i ,-ie^{i\varphi} \right)e^{-\sqrt{M^2_0-\mu_0^2}x}.
\end{equation}
Here,
\begin{equation}
c_1=e^{-i(\frac{\pi}{4}+\frac{\varphi}{2})}\,\,\,;\,\,\,\,\,\,\,e^{i{\varphi}}=\frac{\mu_0-i\sqrt{M_0^2-\mu_0^2}}{M_0}.
\end{equation}
Outside of the gapped region, the solution is obtained by imposing the continuity of the wavefunction. The calculation is trivial because it can be performed component by component thanks to the fact that $\mathcal{H}(x)$ is diagonal for $x<L_S$ and $x>L_D$.\\
The solution is explicitly in the Majorana form. In fact $\alpha(x)=-\delta^*(x)$ and $\beta(x)=\gamma^*(x)$. In the limit $\Delta_0 L_S,\,\,M_0 L_D\rightarrow\infty$, it describes a Majorana bound state. As expected, the decay length in the superconducting region is inversely proportional to $\Delta$, while on the magnetic side it is inversely proportional to $\sqrt{M_0^2-\mu^2}$. In the strong but finite barrier limit, the Majorana bound state is leaking into the gapless contact. 
The existence of a zero energy solution that is peaked around $x=0$ can be understood on general grounds. In fact, it can be shown that the Hamiltonian in Eq.(2) can be mapped onto two disconnected Jackiw-Rebbi models with masses $m_1(x)=\Delta(x)-M(x)$ and $m_2(x)=\Delta(x)+M(x)$. For the  spatial profiles we consider in this article, only $m_1(x)$ has a kink and hence a single zero energy state is present at the interface. 
\subsection{Scattering states at zero energy}
In Sec.3.1, we established that a Majorana bound state solution is present at the interface between the two regions characterized by different gaps, largely independent of the parameters of the system. The zero energy solution is non-degenerate if $L_S$ and $L_D$ are semi-infinite. In fact, in this case, all the solutions that grow away from the interface become non-normalizable, so that the solution in Eq.(3) is the only admissible solution. In order to electrically probe such a Majorana bound state, contacts should be added. The simplest scenario one can consider is a finite barrier and a semi-infinite one, with the gapless region beyond the finite barrier acting as an electric contact. Due to the nature of the Bogoliubov-De Gennes equations, it is possible to built two orthogonal scattering states at zero energy, corresponding to an electron or a hole incoming from the contact. The zero energy subspace becomes two-dimensional. The case of semi-infinite $L_S$ and finite $L_D$ is studied in Ref.\cite{fr}.  Here, the zero energy scattering states are always Andreev backscattered. In this section, we consider finite $L_S$ and $L_D$, so that the zero energy subspace has dimension four. We consider the four scattering states $\psi_S^{(i)}$, $i=1,..,4$, corresponding to an electron incoming from the left ($i=1$), an electron incoming from the right ($i=2$), a hole incoming from the left ($i=3$) and a hole incoming from the right ($i=4$). Explicitly, the scattering states corresponding to an electron ($i=1$) and a hole ($i=3$) incoming from the left, at zero energy, are the eigenstates that, for $x<-L_S$, read
\begin{eqnarray}
\psi_S^{(1)}(x)&=&(e^{i\mu_0 x},r_{e-e}^{(1)}e^{-i\mu_0 x},r_{e-h}^{(1)}e^{i\mu_0 x},0)^T,\\
\psi_S^{(3)}(x)&=&(0,r_{h-e}^{(3)}e^{-i\mu_0 x},r_{h-h}^{(3)}e^{i\mu_0 x},e^{-i\mu_0 x})^T.
\end{eqnarray}
In these equations, the parameters $r_{e-e}^{(i)}$, $r_{e-h}^{(i)}$, $r_{h-e}^{(i)}$, $r_{h-h}^{(i)}$ and the corresponding wavefunctions for $x>-L_S$ are directly determined by solving the Schr\"odinger equation.
For $x>L_D$, we find the general forms for the scattering states
\begin{eqnarray}
\psi_S^{(1)}(x)&=&(t_{e-e}^{(1)}e^{i\mu_0 x},0,0,t_{e-h}^{(1)}e^{-i\mu_0 x})^T,\\
\psi_S^{(3)}(x)&=&(t_{h-e}^{(3)}e^{i\mu_0 x},0,0,t_{h-h}^{(3)}e^{-i\mu_0 x})^T.
\end{eqnarray}
Here, $t_{e-e}^{(1)}$, $t_{h-h}^{(3)}$, $t_{e-h}^{(1)}$ and $t_{h-e}^{(3)}$ are transmission coefficients. Similarly, the scattering states for an electron ($i=2$) and a hole ($i=4$) incoming from the right, are given, for $x>L_D$, by
\begin{eqnarray}
\psi_S^{(2)}(x)&=&(r_{e-e}^{(2)}e^{i\mu_0 x},e^{-i\mu_0 x},0,r_{e-h}^{(2)}e^{-i\mu_0 x})^T,\\
\psi_S^{(4)}(x)&=&(r_{h-e}^{(4)}e^{i\mu_0 x},0,e^{i\mu_0 x},r_{h-h}^{(4)}e^{-i\mu_0 x})^T.
\end{eqnarray}
For $x<-L_S$, we then have
\begin{eqnarray}
\psi_S^{(2)}(x)&=&(0,t_{e-e}^{(2)}e^{-i\mu_0 x},t_{e-h}^{(2)}e^{i\mu_0 x},0)^T,\\
\psi_S^{(4)}(x)&=&(0,t_{h-e}^{(4)}e^{-i\mu_0 x},t_{h-h}^{(4)}e^{i\mu_0 x},0)^T.
\end{eqnarray}
Since we consider strong barriers ($\sqrt{M_0^2L_D^2+\Delta_0^2L_S^2}\gg v_F$, with $v_F$ here restored for clarity), transmission coefficients are usually negligible. The only exception is the parameter region where $M_0^2\approx \mu_0^2+\frac{L_S^2}{L_D^2}\Delta_0^2$ is satisfied. The physical meaning of this condition is discussed in Sec.4.\\
The feature that characterizes all scattering states is the following one. In the strong barrier limit, normal reflection ($|r_{e-e}^{(i)}|^2$ or $|r_{h-h}^{(i)}|^2$) is close to unity whenever 
\begin{equation}
M_0^2> \mu_0^2+\frac{L_S^2}{L_D^2}\Delta_0^2.
\end{equation}
Correspondingly, the local linear conductance $G_0$ satisfies $G_0\sim 0$. On the other hand, when 
\begin{equation}
M_0^2< \mu_0^2+\frac{L_S^2}{L_D^2}\Delta_0^2,
\end{equation}
Andreev reflection ($|r_{h-e}^{(i)}|^2$ or $|r_{e-h}^{(i)}|^2$) is dominant and the linear conductance satisfies $G_0\sim 2e^2/h$.
The paradigmatic case of the coefficients defining $\psi_S^{(1)}(x)$ is shown in Fig.2(a). A density plot illustrating the dependence of the electron-electron reflection on the relevant parameters is presented in Fig.2(b). The black line represents the solution to $M_0^2= \mu_0^2+\frac{L_S^2}{L_D^2}\Delta_0^2$. The abrupt change in the reflection coefficient happens in correspondence with this line in the strong barrier limit, while resonances can produce deviations for small $M_0$ and $\Delta_0$.\\
\begin{figure}
	\resizebox{0.85\columnwidth}{!}{%
		\includegraphics{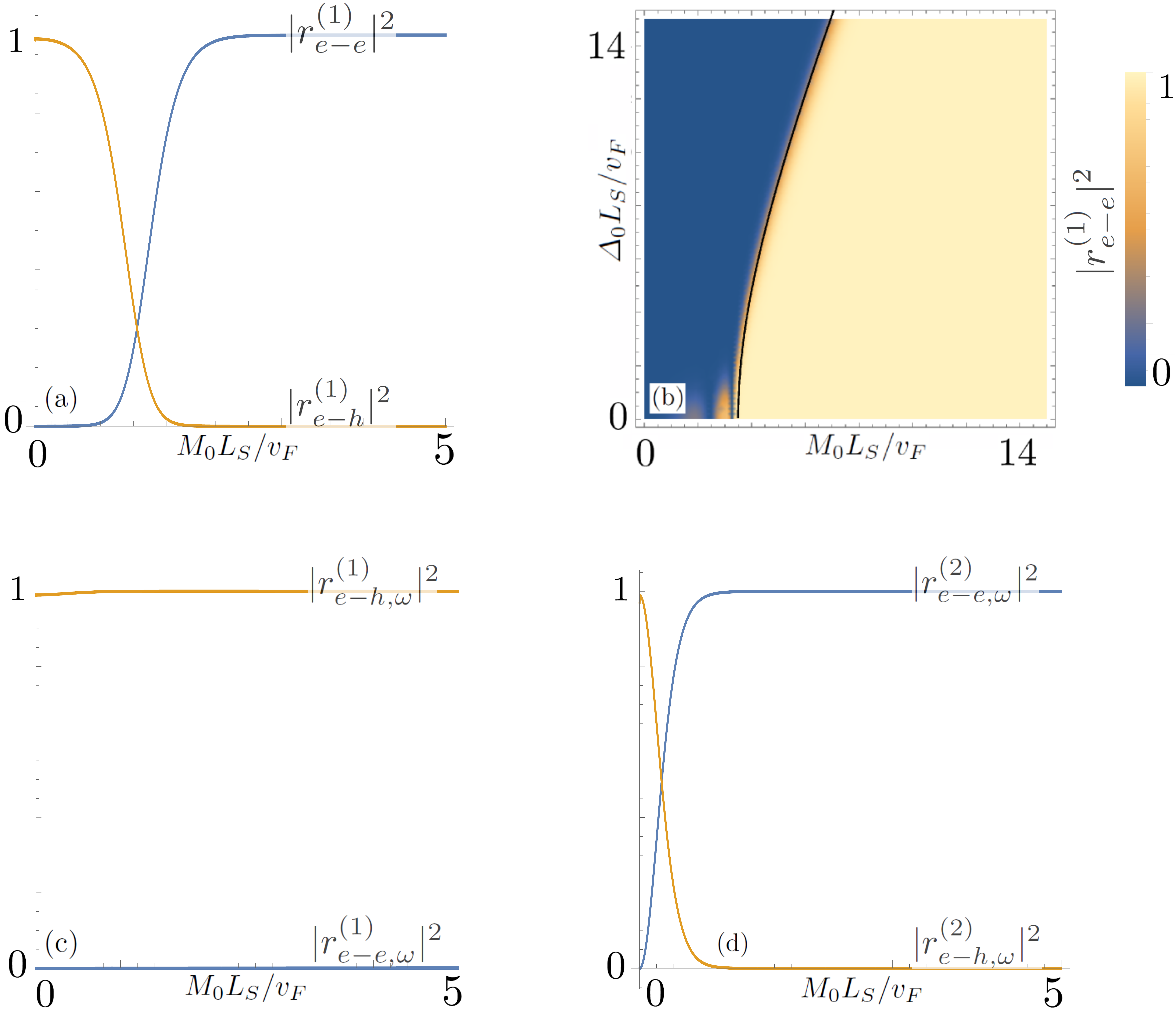} }
	\caption{(a) $|r_{h-e}^{(1)}|^2$ (orange) and $|r_{e-e}^{(1)}|^2$ (blue), as a function of $M_0$, in units of $v_F/L_S$, for $L_D=2.5L_S$, $\Delta_0=3 v_F/L_S$ $\mu_0=0.4 v_F/L_S$. (b) $|r_{e-e}^{(1)}|^2$ as a function of $M_0$ and $\Delta_0$, in units of $v_F/L_S$, for $L_D=2.5 L_S$ and $\mu_0=3.5 v_F/L_S$. The black line is the solution to $M_0^2= \mu_0^2+\frac{L_S^2}{L_D^2}\Delta_0^2$, for the same values of the parameters. (c) $|r_{h-e}^{(1)}|^2$ (orange) and $|r_{e-e}^{(1)}|^2$ (blue), as a function of $M_0$, in units of $v_F/L_S$, for $L_D=2.5L_S$, $\Delta_0=3 v_F/L_S$ $\mu_0=0.4 v_F/L_S$ and $\omega=0.25 v_F/L_S$. (d)$|r_{h-e}^{(2)}|^2$ (orange) and $|r_{e-e}^{(2)}|^2$ (blue), as a function of $M_0$, in units of $v_F/L_S$, for $L_D=2.5L_S$, $\Delta_0=3 v_F/L_S$ $\mu_0=0.4 v_F/L_S$ and $\omega=0.25v_F/L_S$.}
	\label{fig:1}       % Give a unique label
\end{figure}
We have hence shown that, while a Majorana bound state is always present at the interface between the magnetic barrier and the superconductor, the linear conductance of the system shows a sharp transition. The linear conductance is hence not always capable to detect the presence of the bound state, despite the fact that the system is fully coherent.
\section{Explanation of the main result}
The result obtained previously only applies to zero energy. The reason is that, only at zero energy, the spatial ordering of the superconductor and of the magnetic barrier does not influence the transport properties. This implies that the behavior of the scattering states incoming from the left and from the right is the same. A genuine competition of the superconducting and the magnetic gap hence determines all linear conductances. Away from zero energy, the scattering states are mostly influenced by the nature of the first barrier they encounter. Scattering states corresponding to particles incoming from the left show, at low but non-zero energy, mainly Andreev reflection, while the particles incoming from the right are characterized mostly by normal reflection. At finite energy, there is hence no need for a competition between the magnetic and the superconducting gap.\\
In order to better understand this behavior, we start from a rewriting of the Bogoliubov-De Gennes equation at zero energy. We find
\begin{equation}
\partial_x (\alpha(x),\beta(x),\gamma(x),\delta(x))^T=i h(x)(\alpha(x),\beta(x),\gamma(x),\delta(x))^T,
\end{equation}
with
\begin{equation}
h(x)=\mu_0\tau_0\sigma_z-i M(x)\tau_z\sigma_y-i\Delta(x)\tau_y\sigma_z.
\end{equation}
The equation can be formally solved, leading to ($x>x'$)
\begin{equation}
(\alpha(x),\beta(x),\gamma(x),\delta(x))^T= U(x,x')(\alpha(x'),\beta(x'),\gamma(x'),\delta(x'))^T
\end{equation}
with
\begin{equation}
U(x,x')=S_\leftarrow e^{i\int_x^{x'}d\xi h(\xi)},
\end{equation}
and $S_\leftarrow$ the spatial ordering\cite{k1,k2,k3,k4}. Notably, $U(x,x')=U^\dag(x',x)$. The transport properties are encoded in the propagator $U(L_D,-L_S)$. The mathematical reason is that this matrix allows us to calculate the scattering states in the right and left leads, and hence the scattering matrix. Explicitly, the scattering states can be calculated by solving
\begin{equation}
U(L_D,-L_S)\psi^{(i)}_S(-L_S)=\psi^{(i)}_S(L_D),
\end{equation}
corresponding to a system of four equations and four unknowns.\\
To proceed, we now inspect the propagator $U(L_D,-L_S)$. Since the terms in $h(x)$ are stepwise constant, we have
\begin{equation}
U(L_D,-L_S)=e^{iL_D(\mu_0 \tau_0\sigma_z-iM_0\tau_z\sigma_y)}e^{iL_S(\mu_0 \tau_0\sigma_z-i\Delta_0\tau_y\sigma_z)}
\end{equation}
because $[\tau_0\sigma_z,\tau_y\sigma_z]=[\tau_y\sigma_z,\tau_z\sigma_y]=0$. Since multiplying $\psi^{(i)}_S(x)$ by diagonal matrices composed by constant phases does not change the transport properties, the scattering states can be equivalently built by solving the alternative system of equations
\begin{equation}
\tilde{U}(L_D,-L_S)\psi^{(i)}_S(-L_S)=\psi^{(i)}_S(L_D)
\end{equation}
with the effective propagator $\tilde{U}(L_D,-L_S)$ given by
\begin{equation}
\tilde{U}(L_D,-L_S)=e^{i(L_S+L_D)\left(\frac{L_D}{L_S+L_D}(\mu_0\tau_0\sigma_z-i M_0\tau_z\sigma_y)-i \frac{L_S}{L_S+L_D}\Delta_0\tau_y\sigma_z\right)}.
\end{equation}
Remarkably, the same effective propagator would emerge if the position of the magnetic barrier and of the superconductor were exchanged. Hence, at zero energy, the order of the barriers does not influence the transport properties. Interestingly, the effective propagator corresponds to the propagator of a system where the superconductor, the chemical potential, and the magnet, with strengths renormalized by the corresponding lengths, were on top of each other. In particular, $\tilde{U}(L_D,-L_S)$ is the propagator across a quantum spin Hall system gapped, for $-L_S<x<L_D$, by both a superconductor, with effective induced pairing $\tilde{\Delta}={L_S\Delta_0}/({L_S+L_D})$ and a magnetic barrier with $\tilde{M}={L_D M_0}/({L_S+L_D})$. The chemical potential is also renormalized to $\tilde{\mu}={L_D \mu_0}/({L_S+L_D})$. Such a system looks artificial. However, it can be regarded as the small momentum sector $k\sim 0$ of a spin-orbit coupled quantum wire proximitized by a superconductor and subject to a magnetic field parallel to the axis of the wire\cite{loss1,loss2,decaying}. It is known that, in this model, the $k=0$ gap closes and reopens at the topological phase transition, that is for $\tilde{M}^2=\tilde{\Delta}^2+\tilde{\mu}^2$. In terms of the original variables, this relation corresponds to the position of the jump in the conductance. In our case, however, there is no strong connection with the presence of the Majorana state, that is always present.\\
All the derivations above strongly rely on the fact that we consider zero energy. In fact, for finite energy $\omega$, the operator $h(x)$ in Eq.(16) is replaced by $h_\omega(x)=h(x)+i\omega\tau_z\sigma_z$. Since the additional term does not commute with $h(x)$, the spatial ordering in the propagator that could be eliminated in $\tilde{U}(L_D,-L_S)$ remains, spoiling the demonstration provided before. The finite energy scattering states, for which we adopt the names used for the zero energy case, with the addition of an index $\omega$, behave hence as intuitively expected: $\psi^{(1)}_{S,\omega}(x)$ and  $\psi^{(3)}_{S,\omega}(x)$ are dominated by Andreev reflection and result in the local differential conductance $G_L=2e^2/h$ on the left side, at the applied voltage $V=\omega/e$, while $\psi^{(2)}_{S,\omega}(x)$ and $\psi^{(4)}_{S,\omega}(x)$ are dominated by normal reflection and give rise to the local differential conductance $G_R=0$ on the right side. Typical behavior of finite energy scattering is shown in Figs.2(c),(d). Note that the magnetic gap only exists for $M_0<|\mu_0|$. If this condition is not satisfied, there is no magnetic gap and the only relevant gapped region is the superconducting one. This consideration explains the behavior in Fig.2(d), for small $B$.
\section{Conclusion}
We have shown that in a heterostructure consisting of a quantum spin Hall system gapped partially by a superconductor and partially by a ferromagnet, a Majorana bound state is always present at the interface between the two differently gapped regions. However, the linear conductance is not able to capture the presence of the bound state, since, depending on the parameters, it can take both values close to $2e^2/h$ and values close to zero. That is the reason why we have coined it the invisible Majorana bound state. We have demonstrated that the origin of the behavior is that, at zero energy, the spatial ordering appearing in the propagator plays no role in the calculation of the transport properties. We have interpreted the result by mapping the heterostructure onto the low energy model of a proximitized spin-orbit coupled quantum wire in the presence of a magnetic field. Finally, we have shown that, away from zero energy, the spatial ordering becomes again crucial and the expected transport properties are restored.
\section*{Acknowledgments}
We acknowledge financial support by the DFG (SPP1666 and SFB1170 ”ToCoTronics”), the Helmholtz Foundation (VITI), the ENB Graduate school on ”Topological Insulators”, and the Studienstiftung des Deutschen Volkes.

\end{document}